\documentclass[aip,preprint]{revtex4-1}
\draft 
\usepackage{epsf,epsfig}
\begin{document}


\title{Size and density controlled metal nanocluster embedded metal-oxide-semiconductor structure for memory applications} 



\author{D. Biswas}
\email[]{e-mail: debaleen.biswas@saha.ac.in}
\affiliation{Saha Institute of Nuclear Physics, 1/AF Bidhannagar, Kolkata 700 064, India}
\author{S. Mondal}
\affiliation{Maharaja Manindra Chandra College, 20, R.~Bose St, Kolkata 700 003, India}
\author{S. R. Bhattacharyya}
\affiliation{Saha Institute of Nuclear Physics, 1/AF Bidhannagar, Kolkata 700 064, India}
\author{S. Chakraborty}
\email[]{e-mail: supratic.chakraborty@saha.ac.in}
\affiliation{Saha Institute of Nuclear Physics, 1/AF Bidhannagar, Kolkata 700 064, India}

\date{\today}

\begin{abstract}

Metal-nanoclusters (NC), deposited by magnetron-based nanocluster source coupled with quadrupole mass filter (QMF) assembly having independent control over its size and density, are used in fabricating NC-based non-volatile memory (NVM) devices. The effect of diameter and density  on the NVM charge storage characteristics  are presented where Ag is used as the metal NC. The Ag-NC, sandwiched between HfO$_2$ tunnel and control oxides, is deposited by using the combination of the above two instruments. No annealing is performed at any stage of the device fabrication. The largest hysteresis loop area in the capacitance-voltage ($C-V$) characteristics of metal-oxide-semiconductor (MOS) characteristics is observed for a cluster density of 1 $\times$ 10$^{11}$ cm$^{-2}$.  Further, an NC size dependent hysteresis loop area is observed with the MOS devices where the NC diameter is varied from 3 to 1.5 nm keeping the NC density at 1 $\times$ 10$^{11}$ cm$^{-2}$. The device performance is found to be improved with a reduction of the NC size and shows its best with the NC diameter of 1.5 nm.  The storage time of the NVM devices also increases with the decrease in the NC diameter and  exhibits their best performances for the NCs with a diameter of 1.5 nm. 

\end{abstract}


\maketitle 

\section{Introduction}

Nanoparticle (NP)-based nonvolatile memory (NVM) devices have drawn attention of the researchers due to their better storage capacity, high scalability and fast write/erase speed.\cite{lee2011review,zhao2014review,kundu2013gaas,islam2015size,kim2012memory,jeff2011charge,chakraborty2011study,lee2006memory,yun2011room,gupta2012enhancing,lee2005nickel,tang2013charge} In the NVM devices, different nanoparticles e.g. silicon, metals, compounds are used as charge storing nodes that are sandwiched between the tunnel and barrier oxides.\cite{kundu2013gaas,islam2015size,kim2012memory,jeff2011charge,chakraborty2011study,lee2006memory,yun2011room,gupta2012enhancing,lee2005nickel,tang2013charge,tiwari1996silicon,hanafi1996fast} In recent days, high-$\kappa$ dielectric materials are also being employed as barrier and tunnel oxides and an improvement in the memory performance is observed.\cite{lee2005nickel,tang2013charge,yang2007nickel,pei2009memory} Different processes and technological avenues have so far been explored to get the NPs and to control the sizes and densities due to their dependences on the charge storing properties. The major tool employed to deposit the NPs is the sputtering system. The NP diameter and density can easily be varied using this technique. Attempts have also been made to fabricate the NVM devices where the NPs of $\sim$ 4 nm size are grown using molecular beam epitaxy (MBE) and atomic layer deposition (ALD) techniques.\cite{sargentis2005fabrication,mikhelashvili2015optically} But an independent control over the NP diameter and its number density over the surface is difficult in the sputtering and other deposition techniques. Even in tilted target configuration, a variation in the diameter of the NPs is observed.\cite{jeff2011charge} The MBE and ALD techniques also suffer from the similar problem particularly for the NP sizes of $\sim$ 3 nm or less.\cite{sargentis2005fabrication,mikhelashvili2015optically} The study of the NVM-based MOS devices with the NP sizes of $\sim$ 5 nm or less is very important because the NPs with diameter of $<$ 2 nm show the Coulomb blockade effect.\cite{tsuji2004silver,yun2009sub} The NP-based MOS capacitors are found to show best performance with the NC density ranges from $\sim$ 10$^{11}$ to 10$^{10}$ cm$^{-2}$.\cite{jang2009design,verrelli2013nickel} So, a good control over the size and density of the NPs are required to achieve the best device performances. Further, an optimization of these parameters is also required. Very few reports are so far available wherein a simultaneous control over size and density of the NPs is found. This article describes the effect of size of the nanocluster (NC) and their densities on the storage capacities of MOS devices where the NC diameter and density are separately and precisely controlled using magnetron-based nanocluster source equipped with the quadrupole mass filter (QMF) assembly. Silver (Ag) is used here as the metal NC. The device properties namely, memory and charge storage are studied with the diameter and density of the NCs. 

\section{Experimental}


$n$-type Si (100) substrates with resistivity of 0.1 - 0.5 $\Omega-$cm were used to deposit 4 nm-thin HfO$_2$ tunneling oxide on it at 25 $^{\circ}$C by rf magnetron sputtering. After removing the organic, metal impurities and stray oxide, the substrates were immediately loaded into the sputtering chamber. The HfO$_2$ deposition was then carried out in argon plasma at a constant pressure of  $6.6 \times 10^{-3}$ mB using 99.99\% pure HfO$_2$ target and 50 Watt rf power. The grazing incidence x-ray reflectivity (GI-XRR) technique was employed to estimate the film thickness.  

\begin{figure}[]
\centering
\includegraphics[scale=0.29]{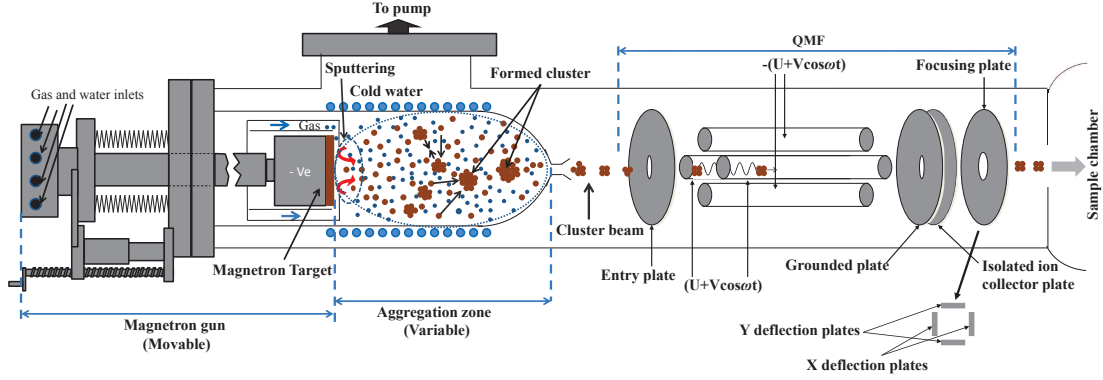}\label{instrument}
\caption{Block diagram of QMF coupled magnetron sputtering-based nanocluster source.}
\end{figure}

A nanocluster deposition unit (Oxford Applied Research; NC200U) was employed to deposit silver nanoclusters (Ag-NC) on already deposited tunnel oxide. The unit consisted of a magnetron-based sputtering system and an aggregation chamber, acted together as  the nanocluster source that was coupled with a quadrupole mass filter (QMF) assembly.\cite{mondal2014performance} A schematic diagram of the nanocluster deposition unit is shown in Fig. 1. The whole unit was utilized to deposit the Ag-NC of the desired density and size on the already deposited HfO$_2$ tunnel oxide on Si.  The magnetron target head was placed on a linear motion drive facilitating the target head to vary the length of the aggregation zone. The distance between the magnetron target head and the aperture before the QMF stage was termed here as aggregation zone length (AZL). The cluster size depended on the AZL and resident time of molecules/atoms in the aggregation region. Argon gas was used here to sputter Ag target for the deposition of the NCs. The aggregation zone was a specially designed water cooled chamber where a temperature of 18~$^{\circ}$C was maintained. The QMF, positioned just after the aperture of the nanocluster source consisted of four cylindrical metal rods that were placed radially as shown in the Fig. 1 and worked following the principle of quadrupole mass spectrometry. A mass spectra for different NCs is obtained by a combination of an ac voltage ($Vcos\omega t$) and dc (U) voltage across the two opposite rods of the QMF assembly. So, two opposite rods experience a potential of $U+Vcos(\omega t)$ and the other two rods have a potential of $-(U+Vcos(\omega t))$. A mass spectrum is obtained by monitoring the ions passing through the quadrupole filter by varying U and V with (U/V) fixed for a constant angular frequency, $\omega$. Depending on the dc and ac voltages applied across the respective quadrupole rods and mass to charge ratio, an ion entering the QMF acquired either a stable or an unstable path.\cite{miller1986quadrupole} If an ac voltage with amplitude V (volts) and frequency f (kHz) was applied, the ions with a particular mass M (amu) passesd through and rest of the ions were blocked. This selected mass M is given by,  

\begin{equation}
M=7\times10^7k\times{\frac{V}{f^2d^2}}
\end{equation}

\noindent where, $k$ is the correction factor which is of the order of unity and $d$ is the diameter of the quadrupole rods that was $\sim$ 25 mm in this case. The radius of the clusters was estimated using the relation $r = r_{ws}n^{1/3}$, where $r_{ws}$ is the Wigner-Seitz radius of the material (for silver, $r_{ws}$ = 1.598 \AA) and n is the number of atoms in the cluster.\cite{lunskens2015plasmons} Molecular/atomic clusters, produced by the dc sputtering, formed a supersaturated admixture of metallic clusters (ionized and neutral). Aggregation of the sputtered molecular/atomic particles took place in the aggregation zone in the presence of Ar and He gases. The unit was so designed that when the pressure in the nanocluster source was $\sim~10^{-1}$ mB, a vacuum of $\sim~10^{-3}$ mB was maintained at the sample chamber. The He gas played a vital role on the cluster diameter in the aggregation zone by quickly reducing the temperature of sputtered atoms/molecules during cluster formation. Ionized clusters, formed with a log-normal size distribution  depended on several source parameters. In the present work, narrow size distributions with peaks for various sizes, separated by $<$ 0.2 nm were obtained by rigorous interplay among the different source parameters  namely, magnetron power, aggregation length, process gas (Ar) and aggregation gas (He). The parameters are presented in {Table- \ref{depo}}. Even after obtaining such a narrow mass distribution, the QMF technique was employed to get the deposition of the desired sized NC with a spread of $\sim$~$\pm4\%$. It is to note that the NC densities of $\sim$ $10^{12}$ cm$^{-2}$ or more produces interference among the NCs resulting in deterioration of the storage performance.\cite{jang2009design,verrelli2013nickel} Therefore, the study on the role of NC density on the storage performance was carried out here considering the maximum density $\sim$~10$^{11}$ cm$^{-2}$. The density was varied independently by varying the deposition time.\cite{mondal2015formation} 

After obtaining a base pressure of 2.0$\times$10$^{-9}$ mB, the Ag sputtering took place at a pressure of 1.8$\times$10$^{-1}$ mB. For each set of deposition, the HfO$_2$ tunnel oxide on Si substrate was mounted in the sample chamber and an amorphous carbon laminated copper grid was placed at its adjacent vicinity for transmission electron microscopy (TEM) studies. The sample chamber pressure was maintained at 1.8$\times$10$^{-3}$ mB. The sample holder was grounded to neutralize the ionized clusters during deposition. The deposited nanoclusters were investigated by a 300 kV TEM facility (FEI TECNAI G2 S-TWIN).

\begin{table}[]
\caption{Variation of Ag cluster size with different nanocluster source parameters.}
\begin{center}
\vspace{1mm}
\label{depo}
\begin{tabular}{ccccc}
\hline
\hline
\vspace{1mm}
Peak at  &  Magnetron & AZL  & Ar flow & He flow \\ 
diameter (nm) &  power (W) & (mm) & (sccm) & (sccm) \\

\hline

1.30 & 11 & 4.8 & 10 & 60 \\
1.50 & 11 & 4.8 & 15 & 60 \\
1.65 & 11 & 14.8 & 15 & 60 \\
1.75 & 11 & 14.8 & 15 & 40 \\
1.82 & 11 & 24.8 & 15 & 60 \\
2.00 & 11 & 54.8 & 15 & 60 \\
2.25 & 26.9 & 34.8 & 15 & 60 \\
2.50 & 26.9 & 104.8 & 15 & 40 \\
2.75 & 40 & 104.8 & 15 & 20 \\
3.00 & 55.6 & 104.8 & 15 & 15 \\
3.25 & 98.7 & 104.8 & 17 & 10 \\  

\hline 
\hline 

\end{tabular}
\end{center}
\end{table}

After the deposition of the NCs, the rf sputtering system was once again employed to deposit a 20 nm-thick HfO$_2$ control gate oxide maintaining the process conditions as that of the tunnel oxide. No annealing was performed on the samples at any stage. For MOS fabrication, aluminium was evaporated on all the samples using electron beam evaporation system. After Al metallization, the gate electrodes of 100 $\mu$m diameter were patterned. The high-frequency capacitance - voltage (HF $C - V$) and capacitance - time ($C - t$) measurements were carried out at 1 MHz using an Agilent E4980A LCR meter. Further $C-t$ measurements were carried out at zero gate bias condition. The gate current - voltage ($I - V$) characteristics were also obtained using Keithley 4200-SCS semiconductor parameter analyzer coupled with pre-amplifiera (4200PA). All the measurements were carried out at room temperature under electrically shielded and light-tight conditions.

\begin{figure}[]
\centering
\includegraphics[scale=0.45]{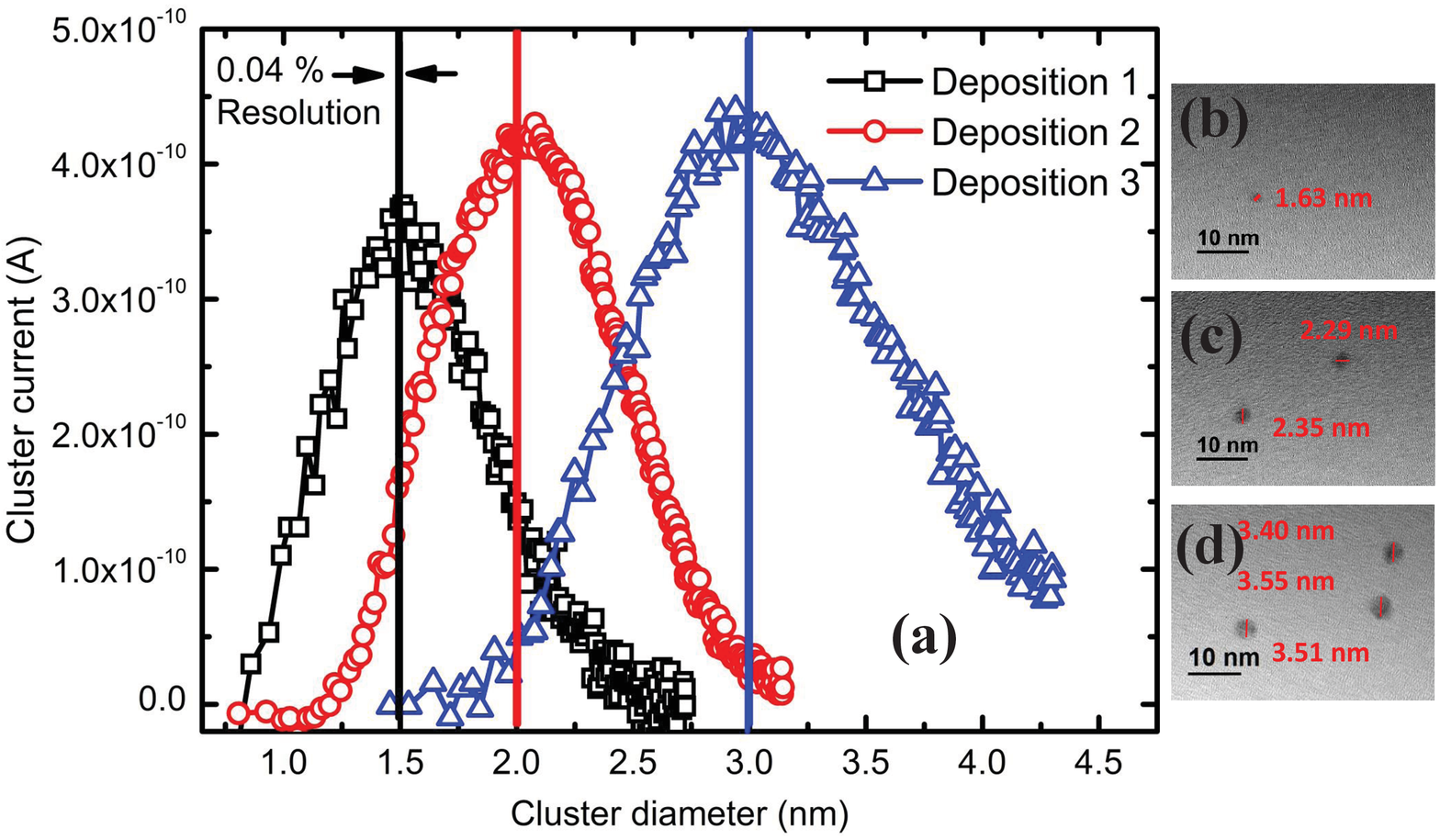}
\caption{(a)Mass spectrum and TEM images of nanocluster of size (b) 1.5, (c) 2.0 and (d) 3.0 nm  }
\label{qmf_tem} 
\end{figure}

\section{Results and Discussions}

A variation of the mass distribution in the form of cluster current is shown in Fig. \ref{qmf_tem}(a) for three deposition conditions to achieve particular NC diameters. The clustering of atoms/molecules in the form of the NCs are mandatory to get an appreciable cluster current at the output of the QMF. The QMF assembly works only if an aggregation of atoms/molecules having a definite width of size distribution, optimized after running the instrument for several times, is achieved. Any cluster diameter can be selected by fixing the corresponding V across the QMF rods and the cluster diameter may be estimated by using Eq. (1). Accordingly, the cluster diameters of 1.5, 2.0 and 3.0 nm are, respectively obtained. The Table. \ref{depo} shows the source parameters for obtaining different cluster sizes ensuring a size-selected mono-dispersed deposition. The representative TEM images obtained from the concurrent deposition on amorphous carbon film, shown in Fig. \ref{qmf_tem}(b)-(d), also corroborate the above observations. It is evident from the TEM image that the clusters gain a finite velocity that in turn changes their shapes after landing over the surface and form a flattened hemispherical islands. As a result, the expected diameter of Ag-NC is found to be more when observed under the TEM. The TEM images show the diameters are  1.6, 2.3 and 3.5 nm instead of 1.5, 2.0 and 3.0 nm, respectively. So, it is clear from the above the Ag-NC clusters are deposited with precise control in size. 

\begin{figure}[]
 \centering
\includegraphics[scale=0.4]{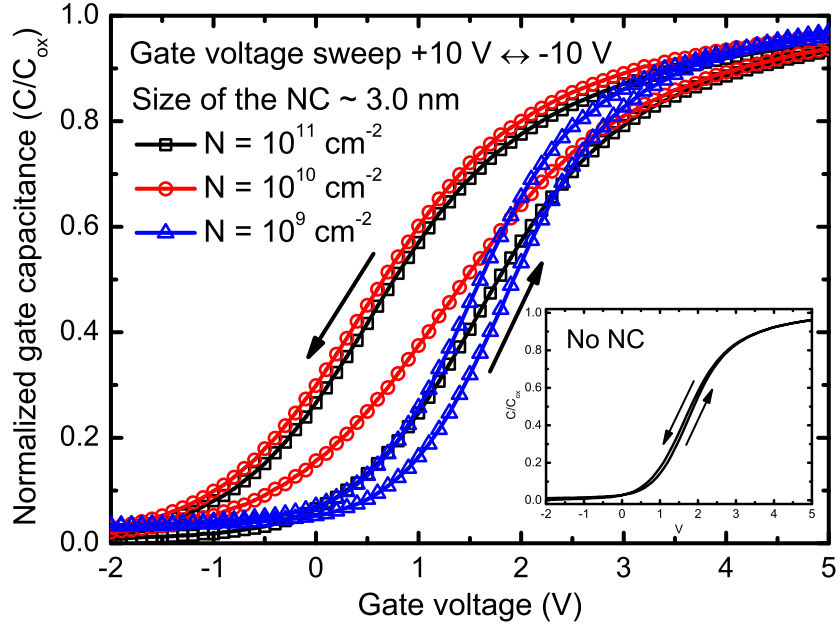}\label{density_cv}
\caption{C-V hysteresis for devices fabricated with different NC density of size 3 nm.}
\end{figure}

\begin{figure}[]
 \centering
 \includegraphics[scale=0.4]{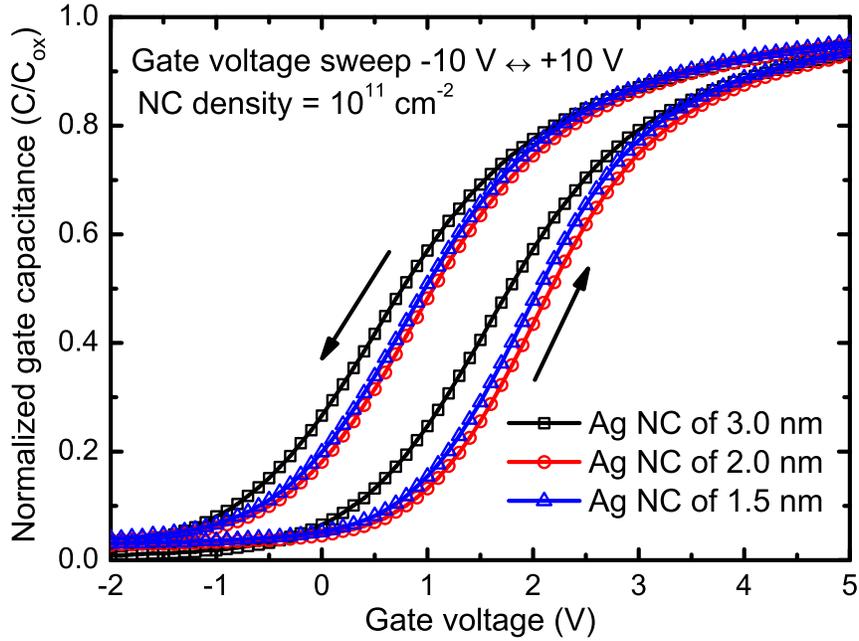}
 \label{size_cv}
\caption{C-V hysteresis for devices fabricated with different NC size of density 1 $\times$ 10$^{11}$/cm$^2$.}
\label{cv}
\end{figure}

The high-frequency $C-V$ plots for 3.0 nm Ag-NC-embedded MOS devices with the different cluster densities are shown in Fig. 3 where the gate voltage sweeps between -10 V $\leftrightarrow$ +10 V. The counter clockwise hysteresis loops in $C-V$ curves indicate hole injection from the substrate to Ag-NCs under negative voltage in the programming cycle and electron injection under positive voltage during the erasing cycle following Fowler-Nordheim (F-N) tunneling.\cite{yun2009sub} The inset of the Fig. 3 shows a negligible hysteresis area for the device without NCs under the same applied voltage sweep indicating the $C-V$ loops are arisen due to the charge storage in Ag-NCs by electron/hole injections. The area under the $C-V$ loop is maximum for the highest nanocluster density i.e. 1 $\times$ 10$^{11}$ cm$^{-2}$ indicating a larger charge storing capacity for the NVM MOS devices. Further, it is already reported that an NC density of $>$ 10$^{12}$ makes the device performance poorer. \cite{jang2009design,verrelli2013nickel} Therefore, it may be construed that an optimum cluster density is required to achieve the best performance of the NVM-based MOS devices which is 1 $\times$ 10$^{11}$ cm$^{-2}$ in this case. Fig. 4 shows a variation of the $C-V$ loops with different sizes of Ag-NCs fixing the cluster density at 1 $\times$ 10$^{11}$ cm$^{-2}$ under the same gate voltage sweep. It is found that the area under the $C-V$ loop is increased when the sizes of the NC decreases from 3.0 nm to 2.0 nm.  But a further reduction of NC size to 1.5 nm does not improve the storage properties significantly.  
 
\begin{figure}[]
 \centering
\includegraphics[scale=0.4]{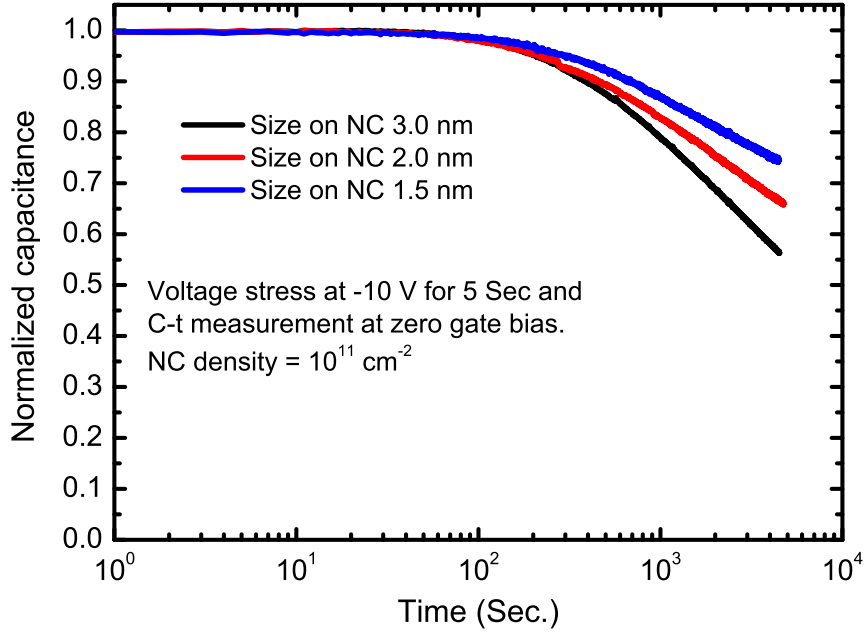}\label{ct_log}
\caption{Variation of C - t for devices fabricated with different NC size of density 1 $\times$ 10$^{11}$/cm$^2$.}
\end{figure}

The charge retention capacity of the Ag-NC-embedded MOS capacitors was studied by measuring the $C-t$ measurements at zero gate bias with different sizes of Ag-NCs. The $C-t$ variations for different NCs are depicted in the Fig. 5.  After charging the Ag-NCs at -10 V for 5 sec, the transient capacitance is measured with no gate bias. The times taken for reduction of charges from its 100\% to 90\% value i.e. the charge retention capacity are 650, 440 and 370 sec for 1.5, 2.0 and 3.0 nm-based NVM devices, respectively.  The charge retention capacity is found to be better for the devices with smaller size of the Ag-NCs probably due to the Coulomb blockade effect below sub 2 nm regime.\cite{kundu2013gaas,yun2009sub} Therefore, smaller size floating NCs with optimum cluster density are preferred for non-volatile memory cell applications.

\section{Conclusion}

The dependence of size and density of Ag-NC on the charge storage characteristics of the NVM-based MOS devices are presented. The largest hysteresis loop area in the C-V characteristics is observed for a cluster density of 1 $\times$ 10$^{11}$ cm$^{-2}$.  Further, largest hysteresis loop area is found for the NVM devices with the NC diameter of 1.5 nm where the cluster density is fixed at 1 $\times$ 10$^{11}$ cm$^{-2}$. The device performance has been found to be improved with a reduction of the NC size. The storage time of MOS devices also increases with the decrease in the NC diameter and performs its best for NCs with a diameter of 1.5 nm. 
 
\section*{Acknowledgment}

The authors wish to acknowledge the contribution of Dr. B. Satpati amd Dr. M. K Mukhopadhyay of the Saha Institute of Nuclear Physics (SINP) for providing the TEM and x-ray facilities. The contributions of Mr. D. Dey of SINP and Mr. N. Alam of Citizen Enterprise are also remembered. Finally, the authors wish to acknowledge the Department of Atomic Energy (DAE), Government of India for financial assistance.


\bibliography{nvm}

\end{document}